\documentclass{aa} 
\usepackage{graphicx}
\usepackage{amsmath} 
\usepackage[utf8]{inputenc}
\usepackage{hyperref}
\newcommand{\upsub}[1]{\sb{\mathrm{#1}}}
\begingroup\lccode`~=`_\lowercase{\endgroup\let~\upsub}
\makeatletter
\newcommand\footnoteref[1]{\protected@xdef\@thefnmark{\ref{#1}}\@footnotemark}
\makeatother

\AtBeginDocument{%
  \catcode`_=12
  \mathcode`_="8000
}
\usepackage{txfonts}
\usepackage{natbib}
\usepackage{footmisc}
\bibpunct{(}{)}{,}{a}{}{,}
\newcommand{\kmps}{km\,s$^{-1}$}
\newcommand{\kmpsb}{km\,s$^{-1}$ }

\newcommand{\ndhp}{N$_2$H$^+$}
\newcommand{\ndhpb}{N$_2$H$^+$ }
\newcommand{\nddp}{N$_2$D$^+$}
\newcommand{\nddpb}{N$_2$D$^+$ }
\newcommand{\hdb}{H$_2$ }

\newcommand{\htp}{H$_3^+$}
\newcommand{\htpb}{H$_3^+$ }
\newcommand{\dtp}{D$_3^+$}
\newcommand{\dtpb}{D$_3^+$ }
\newcommand{\hddp}{H$_2$D$^+$}
\newcommand{\hddpb}{H$_2$D$^+$ }
\newcommand{\ddhp}{D$_2$H$^+$}
\newcommand{\ddhpb}{D$_2$H$^+$ }
\newcommand{\dcop}{DCO$^+$}

\newcommand{\cdsb}{column densities }

\newcommand{\sqc}{cm$^{-2}$}
\newcommand{\sqcb}{cm$^{-2}$ }

\newcommand{\jqt}{(\emph{J}:4 -- 3)}
\newcommand{\jqtb}{(\emph{J}:4 -- 3) }

\newcommand{\jkk}{\emph{J}$_\mathrm{KK\arcmin}$}

\newcommand{\mjy}{MJy\,sr$^{-1}$}

\newcommand{\Av}{A$_{\mathrm V}$}
\newcommand{\Avb}{A$_{\mathrm V}$ }

\newcommand{\pdix}[1]{$\times$\,10$^{#1}$}
\newcommand{\pdixb}[1]{$\times$\,10$^{#1}$\,}

\newcommand{\down}[1]{\textsubscript{#1}}
\newcommand{\mum}{$\mu$m}

\begin{document}

   \title{First map of \ddhpb emission revealing the true centre of a prestellar core: further insights into deuterium chemistry} 
   \subtitle{} 

   \author{L. Pagani           \inst{1}
          \and
          A. Belloche\inst{2}
          \and
          B. Parise \inst{2} \thanks{The affiliation of B. Parise corresponds to her affiliation when the project was initiated.}            
         }

   \offprints{L.Pagani}

 \institute{ LERMA \& UMR8112 du CNRS, Observatoire de Paris, PSL Research University, CNRS, Sorbonne Universités,  F- 75014 Paris, 
 France\\
\email{laurent.pagani@obspm.fr}
\and
 Max-Planck-Institut f\"ur Radioastronomie, Auf dem H\"ugel 69, 53121 Bonn, Germany
           }

   \date{Received 04/07/2023; accepted ...}

 
  \abstract
   {\object{IRAS 16293E} is a rare case of a prestellar core being subjected to the effects
 of at least one outflow.}
   {We want to disentangle the actual structure of the core from the outflow impact and evaluate the evolutionary stage of the core.}
   {Prestellar cores being cold and depleted, the best tracers of their central regions are the two  isotopologues of  trihydrogren cation which 
   are observable from the ground, ortho--\hddpb and para--\ddhp. We used the Atacama
   Pathfinder EXperiment (APEX) telescope to map the para--\ddhpb emission in
   \object{IRAS 16293E} and collected James Clerk Maxwell Telescope (JCMT) archival data of ortho--\hddp.
    {We compare their emission to that of other tracers, including dust emission, and analyse their abundance with the help of a 1D radiative transfer tool. 
       The ratio of the abundances of ortho--\hddpb and para--\ddhpb can be used to estimate the stage of the chemical evolution of the core.}}  
   {We have obtained the first complete map of para--\ddhp\ emission in a prestellar  core. We compare it to a map of ortho--\hddp and show their partial
    anti-correlation. This reveals a strongly evolved core with a para--\ddhp/ortho--\hddpb 
    {abundance ratio towards the centre for which we obtain a conservative lower limit  from 3.9 (at 12\,K) up  to 8.3 (at 8\,K) while the high extinction of the core 
    is indicative of a central temperature below 10\,K.}
    {This ratio is higher than predicted by the known chemical models found in the literature.}
   Para--\ddhpb (and indirectly ortho--\hddp) is the only species that reveals the true centre of this core, while
   the emission of other molecular tracers and dust are biased by the temperature
   structure that results from the impact of the outflow.}
   {This study invites to reconsider the analysis of previous observations of this source and
   possibly questions the validity of the deuteration chemical models 
   or of the reaction and inelastic collisional rate coefficients of the \htpb isotopologue family. {This could impact the deuteration clock predictions for all sources}.}


\titlerunning{IRAS 16293E: first map of \ddhpb }

\maketitle
%

\section{Introduction}
\object{IRAS 16293E} is a well-known prestellar core (PSC) at a distance of 141 pc \citep{dzib_revised_2018} which is under the influence of at least one outflow coming from the 
multiple protostar 
system 
\object{IRAS 
16293--2422} \citep[e.g.][]{Wootten1987, Castets2001, Lis2002,  Stark:2004hz}. A second outflow, visible in the Spitzer/IRAC data and coming from the Young 
Stellar 
Object (YSO) \object{WLY 2-69} was also reported by \citet{pagani_iras16293e_2016}. This outflow points towards the south of the 
PSC, close to 
the HE2 
position \citep{Castets2001} but whether this represents a fortuitous alignment or reveals
an actual impact is not yet established. \object{IRAS 
16293--2422} has attracted considerable attention as well, with the SMA 
\citep{jorgensen_arcsecond_2011} and ALMA \citep[program PILS, ][ and follow-up papers]{jorgensen_alma_2016}, and the first (and still 
unique) 
detections of para--\hddpb and ortho--\ddhpb in absorption in front of the Class 0 objects \citep{Brunken:2014uc,harju_detection_2017}.
\object{IRAS 16293E} has also gained new attention with the advent of ALMA observations combined with CSO maps \citep{Lis:2016gx}
after the detection of ND$_3$ \citep{Roueff:2005jh}, which confirm that most deuterated species such as \nddpb or DCO$^+$  
peak at the outflow-PSC interface, with the exception of ND$_2$H and ND$_3$, the emission of which peak about 10\arcsec\ east of the DCO$^+$ emission peak.
In a preliminary study we showed that Herschel continuum observations of the source with PACS and SPIRE  reveal the temperature gradient induced by the 
outflow 
while  ortho--\hddpb emission traces the cold part of the core, away from the working  surface, and approximately coincidental with the core seen in absorption in 
Spitzer IRAC mid-infrared images  \citep{pagani_iras16293e_2016}. 
The recent study by \citet{kahle_molecular_2023} of a large set of observations
performed with the Atacama Pathfinder EXperiment (APEX) telescope towards both
the YSOs and the PSC improves the census of molecular abundances and better
differentiates the various parts of this complex region, which confirms our
understanding of its structure.

{It has been shown that the deuterium chemistry evolves in a rather monotonic way during the PSC formation phase and measuring the abundance of deuterated species 
could 
help to measure the time the cloud took to contract to the PSC core under scrutiny
\citep{Flower:2006ea, Pagani2009a, pagani_ortho-h2_2013}. Though it is easier to measure deuteration of species like NH$_3$, HCO$^+$, or \ndhp, it has been advocated that 
measuring directly the 
abundance of the deuteration vector, namely the \htpb isotopologues, would be more precise to characterize the evolution of the deuteration \citep{parise_extended_2011, 
harju_detection_2017}. However single point 
measurements are often ambiguous, provide only line of sight averaged values, often decreasing the contrast between species. Only maps of the species allow us to fully understand 
the physics and chemistry of the PSCs.}

In this paper, we present the first  map of para--\ddhp, towards the prestellar core \object{IRAS 16293E}, and compare it with archival data of 
ortho--\hddp, dust emission and \ndhp. Section 2 presents the observations.
{In 
Sect.~3, we analyse the emission of both deuterated isotopologues of \htp, and compare them to dust absorption and emission, and \ndhpb emission. 
We present a 1D modified Monte-Carlo model in Sect. 4 to quantify the abundance of both isotopologues,  and discuss the various results in Sect. 5.}
Our conclusions are given in Sect. 6.
%

\begin{figure*}[t!]
\centering
\includegraphics[width=18cm]{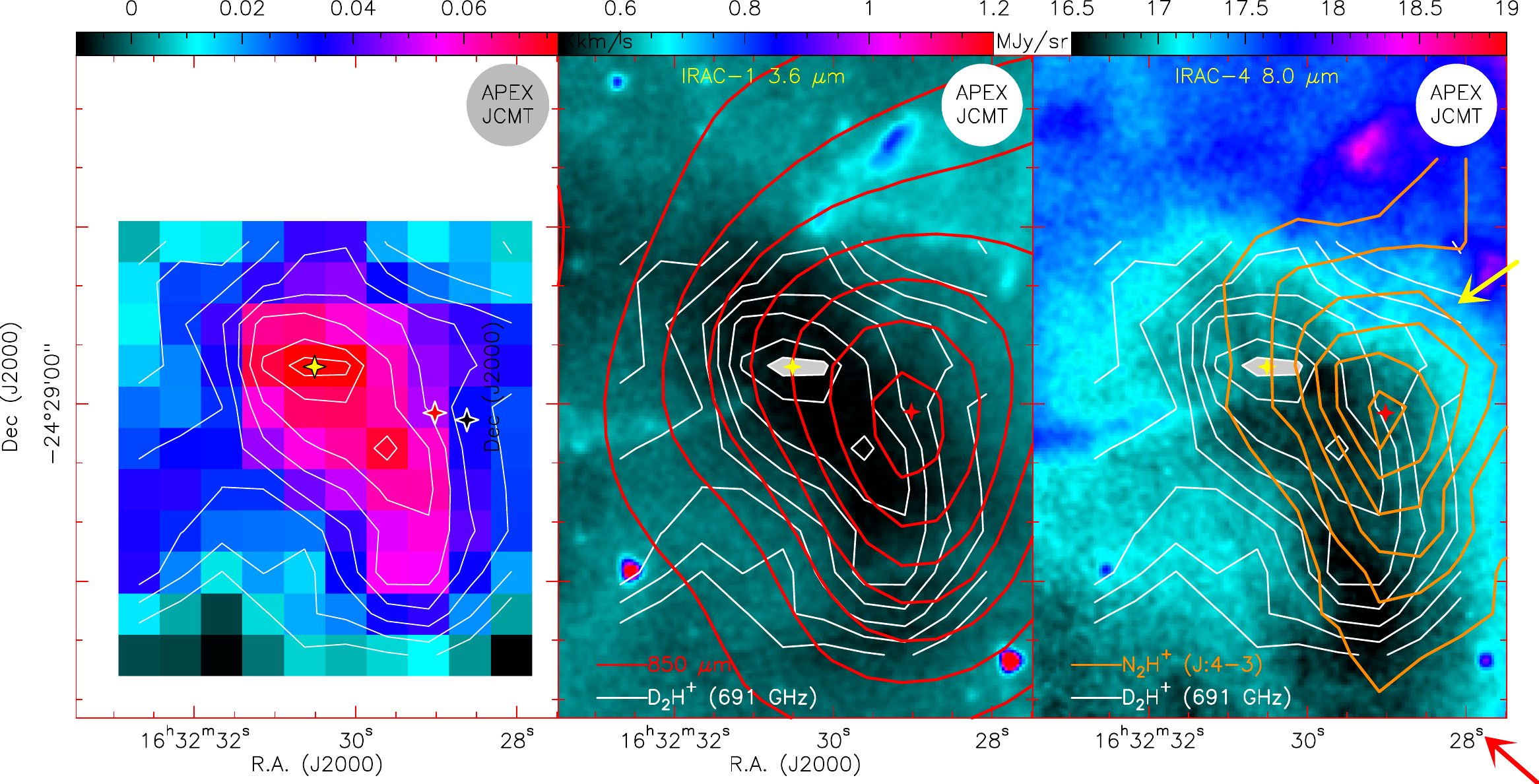}
\caption{Continuum and line emission images of \object{IRAS 16293E}. {\bf Left:} Colour map of para--\ddhpb 
integrated intensity emission with white contours (colour scale and white contours from 0.01 to 0.07 by 0.01\,K\,\kmps plus 
a contour at 0.075\,K\,\kmps ; contour spacing of 0.01\,K\,\kmps\,= 1.5$\sigma$)
{\bf Centre:} IRAC-1 image at 3.6\,\mum. White contours trace para--\ddhpb as in the left panel. Red contours trace the SCUBA-II 850\,\mum\ emission (20 to 160 by 20 
[$\approx$\,10$\sigma$]\,\mjy).
{\bf Right:} IRAC-4 image at 
8.0\,\mum. White contours trace para--\ddhpb as in the left panel. Orange contours trace the \ndhp\,\jqtb integrated 
intensity (0.3 to 2.3 by 0.4 [=\,12$\sigma$] 
K\,\kmps). The yellow cross marks the centre of the \ddhpb emission ($\alpha_{J2000}$:\,16$^h$32$^m$30.51$^s$, 
  $\delta_{J2000}$: \,$-$24$\degr$28$^\prime$53.7$^{\prime\prime}$, see also Fig. \ref{fig:figure2H2DpD2Hp}). The red cross marks the peak 
position of the dust and \ndhp\,\jqtb 
emissions.  The black cross (left panel) marks the position of the observation of \ddhpb and \hddpb by \citet{vastel_detection_2004}.
{The yellow arrow traces the direction of the outflow from the IRAS16293--2422 protostellar system and the red arrow the outflow from WLY 2-69}. The grey and white disks 
represent the JCMT and APEX beams after smoothing to 14\arcsec.}
\label{fig:figure1irac-14d2hp850mun2hp43}
\end{figure*}

\section{Observations}
We  observed para--\ddhp (J$_{K_aK_c}$: $1_{10}$--$1_{01}$) at 691.6604434 GHz \citep{jusko_double_2017} with the Swedish-ESO PI Instrument for APEX (SEPIA660) 
receiver on 
the  
APEX 12-m 
telescope equipped with Fast Fourier Transform (FFT) backends, each of 65536 channels of 61 kHz width. The observations were performed in several runs on ESO 
time (3/5/2021, 28-29/6/2021, project E-0105.C-0251A) 
and MPI time (24-26/3/2022, 1/4/2022, 4/4/2022, project M-0109.F-9501C-2022). We also included observations with the 7 pixel CHAMP+ camera taken on 
7/9/2010 (one pointing, project M-085.F-0016-2010).
The weather was exceptionally good (precipitable water vapour, PWV = 0.2--0.3 mm) during most of the SEPIA660 observations, and very good (PWV $\approx$ 0.5 mm) for the 
CHAMP+ 
observations.
Apart from the seven positions observed with CHAMP+ and two positions observed with SEPIA660 in May 2021, which were observed in position switching mode,
 all observations were done in the OTF 
mode to cover a 50\arcsec\ 
$\times$ 50\arcsec\ region (June 2021) extended to 50\arcsec$\times$72\arcsec\ in 2022 as the preliminary results indicated an unexpected emission to the south.
{The various reference positions are summarized in Table \ref{tab:refpos}.}
The total observing time is about 25 hours. The angular resolution is $\sim$9\arcsec. The pointing was checked every hour, mainly 
on the nearby object RAFGL1922. The stability of the pointing is better than 3\arcsec.
{We used CLASS\footnote{\url{http://www.iram.fr/IRAMFR/GILDAS}} to subtract first-order baselines computed on the velocity range $-$24 to $+$31 \kmps,    and smoothed 
the 
data
(spatially, using XY\_MAP) to 14\arcsec\ }and (spectrally) to 244 kHz
 ($\approx$ 0.1\,\kmps) to improve the sensitivity. {The peak signal-to-noise ratio (SNR) is 7 in brightness temperature and 12 in integrated intensity (in the range 2.81--4.18 
 \kmps). 
 The rms noise level in integrated 
 intensity is  6.7 mK\,\kmpsb  (T$_a^*$ scale) towards the centre of the map and reaches $\approx$ 11 mK\,\kmpsb  at the edges.}
{ We assumed a main beam efficiency of $\eta_{MB}$ $\approx$ 0.48\,$\pm$\,0.048 for CHAMP+ observations\footnote{ 
 https://www.mpifr-bonn.mpg.de/4480868/efficiencies} and $\eta_{MB}$ $\approx$ 0.46\,$\pm$\,0.046 for SEPIA660 observations (Perez-Beaupuits, priv. comm.) which 
 were 
 applied to each set before merging them.}
\begin{figure}[t]
\flushright
\includegraphics[width=\linewidth]{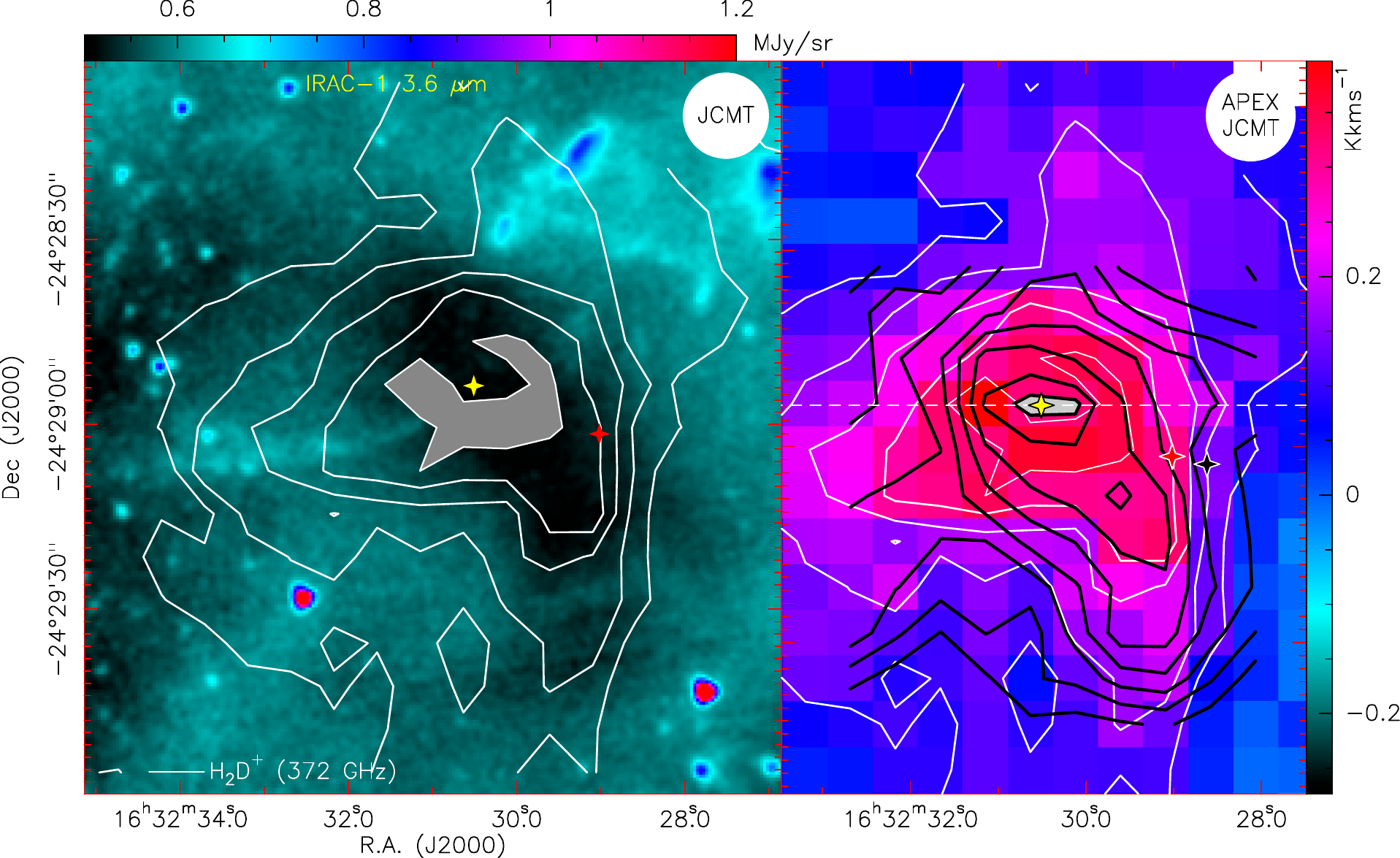}
\caption{{\bf Left:} IRAC-1 image at 3.6\,\mum. White contours trace ortho--\hddp.
The filled grey contour marks the strongest emission (contours from {0.10 to 0.34 }by 0.06  [=\,2.6$\sigma$]\,K\,\kmps). 
{\bf Right:} {para--\ddhpb and ortho--\hddpb maps. Same map of para--\ddhpb (black contours) as in Fig. \ref{fig:figure1irac-14d2hp850mun2hp43} superimposed on the 
ortho--\hddpb 
map (colour 
scale, plus white contours as in left panel). The yellow, black, and red crosses are as in Fig. \ref{fig:figure1irac-14d2hp850mun2hp43}.
 \ddhp, \hddp, and \ndhp spectra along  the cut traced by the dashed line are displayed in Fig. \ref{fig:figure3-1d2hph2dpfit}}}
\label{fig:figure2H2DpD2Hp}
\end{figure}

 We retrieved two sets of unpublished ortho--\hddpb (J$_{K_aK_c}$: $1_{10}$--$1_{11}$) observations at 372.421340 GHz \citep{jusko_double_2017} from the James 
 Clerk Maxwell 
 15-m Telescope (JCMT) 
 archive\footnote{\url{https://www.cadc-ccda.hia-iha.nrc-cnrc.gc.ca/}}, {from projects M07AU29 (2007) and M09AU01 (2009)}.
 They were obtained with the 16 pixel Heterodyne Array Receiver Program (HARP)  receiver with the Auto Correlation Spectral Imaging System (ACSIS) backend 
 (61 kHz sampling). {The atmospheric opacity at 225 GHz ranged from 0.052 to 0.034
  (M07) and from 0.04 to 0.02 (M09). The M07 data consists of a 600-s long
  single pointing of 15 pixels (one corner detector was dead) towards the core
  centre. The M09 data (14 effective pixels) cover 232\arcsec$\times$113\arcsec, one square
  jiggle map centred on the core (1200 seconds integration) with an adjacent jiggle map to
  the East (300 seconds integration). The M09 central map} was presented in our conference poster 
  \citep{pagani_iras16293e_2016}. 
The original angular resolution { of 13.2\arcsec} was smoothed to 14\arcsec\ (after {third-order baseline subtraction on a baseline extending from $-$21 to $+$28\,\kmps, 
resampling
and smoothing} with CLASS/XY\_MAP). The frequency sampling has been Hanning-smoothed to 122 kHz 
($\approx$ 0.1 \kmps).
{We assumed a main beam efficiency of $\eta_{MB}$ $\approx$ 
0.6\footnote{https://www.eaobservatory.org/jcmt/instrumentation/heterodyne/harp/}.
The peak SNR is 11 in brightness temperature and 19 in integrated intensity (in the range 2.92--4.28 \kmps). The average integrated 
intensity rms noise is 23 mK\,\kmpsb for the central map, 
twice as 
high for the eastern map of which we only use the edge to reach the emission limit of the core.}
The same observing run also provided us with \ndhpb \jqtb observations at 372.6734633 GHz \citep{cazzoli_precise_2012}, 
representative of many of the species observed towards the PSC 
\citep{Lis:2016gx,kahle_molecular_2023}. {Both \ddhpb and \hddpb data sets are resampled on a 7\arcsec\ basis.}

We also used the JCMT SCUBA-II 850 \mum\ continuum observations obtained and published by \citet{Pattle:2015bq}. We refer the reader to that paper for a detailed 
description of the 
observations. 
We retrieved Spitzer IRAC data from the archive\footnote{\url{https://sha.ipac.caltech.edu/applications/Spitzer/SHA/}, 
aorkeys : 3651840, 5756160, 18324736, and 47121664}. As far as we know, the PSC image from these observations has never been published 
apart 
from 
 our conference poster \citep{pagani_iras16293e_2016}.

\section{Analysis}\label{Analyse}

Figure \ref{fig:figure1irac-14d2hp850mun2hp43} shows the map of para--\ddhpb  (left panel). This is the first time that a map of para--\ddhpb is presented. 
The doubly deuterated ion follows precisely the IRAC--1 absorption image seen at 3.6 \mum\ (central panel). Since \ddhpb is expected to peak in the central coldest part of clouds, 
and since the total 
absence of 
stars at 3.6 \mum\ reveals high extinction (typically \Avb $\geq$ 100 mag), these two maps do trace the centre of the \object{IRAS16293E} PSC, while the dust emission map 
peak (red cross), which is very sensitive to the temperature increase due to the outflow collision with the PSC edge, is displaced towards the impact point. Therefore 
the dust emission map does not reveal the true cold centre of the PSC. In the right panel, 
{the upper part of the IRAC-4 image of the PSC also seen in absorption is partially hidden (reduced contrast)  by PAH emission in relation with another outflow mostly seen in the 
IRAC images. In the same panel, \ndhpb \jqtb contours (in orange) also point towards the impact point, known as the DCO$^+$ peak \citep[e.g.][]{Lis:2016gx}}.

The left panel of Fig. \ref{fig:figure2H2DpD2Hp} shows the ortho--\hddpb emission superimposed on  the IRAC--1 image. It is again clearly shifted to the East compared to the dust 
emission peak
and 
\ndhpb \jqtb
line emission  peak (marked by the red cross) as we already reported in \citet{pagani_iras16293e_2016} and has been confirmed by \citet{kahle_molecular_2023}. 
It is remarkable that both dust and \ndhpb emissions peak at the same place, probably revealing the exact position of the outflow impact
on the PSC. On the other hand, like for para--\ddhpb emission, dust absorption as seen by IRAC images, is correlated with ortho--\hddpb emission. 
Dust in absorption is not sensitive to temperature effects and this explains the decorrelation between dust absorption and emission. The dust absorption reveals
the highest extinction part of the PSC and as expected this coincides with the emission peak of ortho--\hddpb and para--\ddhp.
Another noticeable feature is the {horse--shoe shape} of the ortho--\hddpb emission surrounding the PSC centre
 (the topmost grey-filled contour in Fig. \ref{fig:figure2H2DpD2Hp}, left side).
This ortho--\hddpb horse--shoe shape clearly surrounds the  emission peak of para--\ddhpb as shown in the right panel of Fig. \ref{fig:figure2H2DpD2Hp}.
This indicates that the ortho--\hddpb emission is weaker towards the centre of the PSC. Since it is only the top contour of ortho--\hddpb emission
which shows this feature, it means that only the inner part of the PSC  shows a diminished emission and therefore a diminished ortho--\hddpb abundance, compared to the outer 
layers.
The intensity difference between the ortho--\hddp peak emission at a right ascension offset of +7\arcsec\ from the centre and the central position is
66 mK\,\kmps, with a statistical noise of 29 and 21 mK\,\kmps, respectively. The difference is  2.3\,$\sigma$, but if the ortho--\hddpb abundance had been 
constant, the emission would have increased towards the centre, like for para--\ddhp, instead of decreasing. Therefore  the difference with respect to a situation where the 
abundance would be constant is much higher than 2.3\,$\sigma$ and statistically 
significant. 
In Sect. \ref{sub:paraortho}, we will model the case of a constant ortho--\hddpb abundance to quantify the expected difference.

The pointed observations of ortho--\hddpb and para--\ddhpb that were previously reported by \citet{vastel_detection_2004} towards the position marked by a black cross in Figs. 
\ref{fig:figure1irac-14d2hp850mun2hp43} \& \ref{fig:figure2H2DpD2Hp} are found to be inconsistent with our more sensitive mapping observations towards the same position.  
For para--\ddhp, they 
reported
a peak 
temperature of 0.34 K  ($\pm$0.077 
K ?)\footnote{\label{note6}In their paper, no numerical value of the rms error is provided, this value is an educated guess from their Fig. 3} and a 
linewidth of 0.29 $\pm$0.07\,\kmps, to be compared with our 83\,$\pm$17\,mK and a linewidth of 0.48\,$\pm$0.07\,\kmpsb; for ortho--\hddp, they report a
peak temperature of 1.31 K ($\pm$0.22 K ?)\footnotemark[6 
] and a linewidth of 0.36\,$\pm$0.04\,\kmps, to be compared with our 0.31\,$\pm$0.03\,K and a linewidth 
of 
0.61\,$\pm$0.04\,\kmps. For a 
gas temperature in the range 16--25\,K from former studies \citep{Loinard:2001im,Castets2001,Lis2002,Stark:2004hz,Lis:2016gx} that applies to the DCO$^+$ emission peak 
region, the thermal linewidth alone is expected to be between 0.38 and 0.48\,\kmpsb for para--\ddhpb and between 0.43 and 0.54\,\kmpsb for ortho--\hddp, which are both 
inconsistent with the small linewidths
reported by \citet{vastel_detection_2004}.
 \begin{figure*}
\includegraphics[angle=90,width=0.96\textwidth]{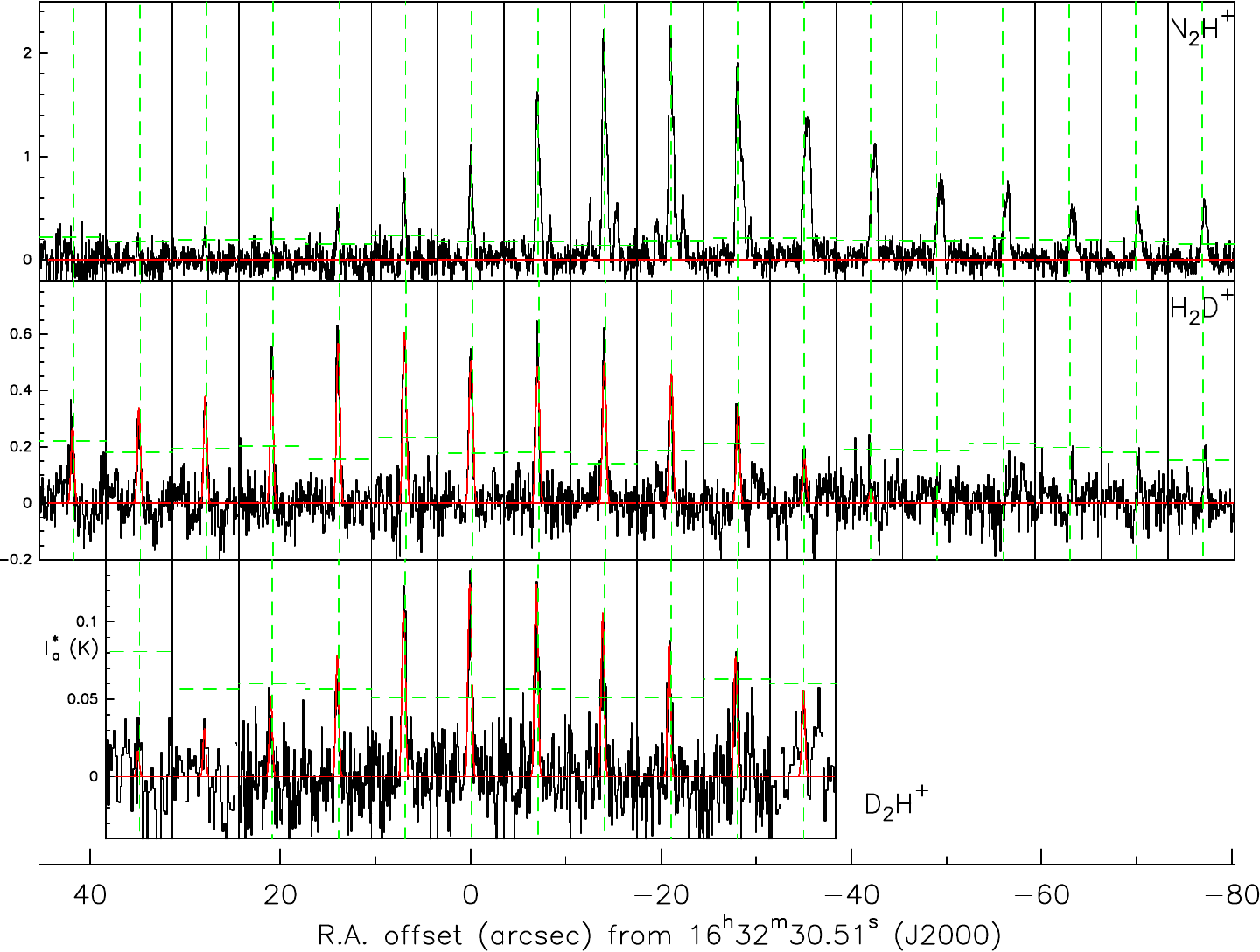}
 \caption{{From top to bottom: \ndhpb\jqt, ortho--\hddp\, (J$_{K_aK_c}$: $1_{10}$--$1_{11}$) and para--\ddhpb (J$_{K_aK_c}$: $1_{10}$--$1_{01}$) spectra in black, 
 extracted along the cut displayed in Fig. \ref{fig:figure2H2DpD2Hp}, with best-fit model of ortho--\hddpb and para--\ddhpb overlaid in red.
 The three weakest para--\ddhpb spectra have been smoothed to 0.21\,\kmps. The velocity axis in each box runs from 1 
 to 6 \kmps. Green dashed lines mark the 3$\sigma$ individual noise levels and the systemic velocity (3.6 \kmps). }}
 \label{fig:figure3-1d2hph2dpfit}
 \end{figure*}

 \section{Radiative transfer modelling of para--\ddhpb and ortho--\hddpb}\label{sub:paraortho}
Modelling the abundance of both \htpb isotopologues is needed to quantify our findings. The shape of the PSC is not symmetrical but still relatively simple. 
There is 
however no easy way to derive its density and temperature structures, due to the complexity of its 
environment. Star counts or reddening measurements are impossible towards the centre of the core due to the lack of stars even in the sensitive Spitzer 
data
 (Fig. \ref{fig:figure1irac-14d2hp850mun2hp43}). 8 \mum\ continuum absorption measurements \citep{Bacmann2000, Lefevre:2016hr}  are hampered by the 
 superposition of nebulosity emission from the 
\object{IRAS 16293--2422A/B} outflows and possibly some Polycyclic Aromatic Hydrocarbon (PAH) emission. All emissions, whether dust or molecular, are 
strongly biased by the 
temperature gradient induced by the outflow impact on the PSC, and the density profile is also possibly perturbed on that side, away from a simple Plummer-like profile 
\citep{whitworth_empirical_2001}.  To disentangle density, 
temperature, and abundances for molecules, or dust properties for continuum emission, a thorough analysis in 3D of the core must be performed. This is beyond the 
scope of 
the present paper and will be addressed in the future. Here, we present a  model combining three partially different one-dimensional
(1D) spherical models.

The first step is to build {1D }gas density and temperature structures of the PSC.  The only tracer from the edge to the centre of the PSC is  dust, especially 
in emission around 300 GHz but its
properties are not accurately known, and since the temperature and density profiles cannot be easily retrieved from 
the dust emission alone \citep{pagani_can_2015}, {we have constructed the core structure from the following considerations:
\begin{itemize}
\item except  for the south-western extension, the emission of both para--\ddhpb and ortho--\hddpb is relatively round and symmetrical. We consider that the bulk of the original  
core, before the outflow impacts  its western side, is still present and close enough to a spherical object.
\item The eastern part of the core is unperturbed and its \hdb density can be described with a Plummer{-like} profile \citep{Tafalla2002}: 
 \begin{equation}\label{plummer}
 \mathrm{n(H_2)(r) = \frac{1.5\times10^6}{1+(\frac{r}{3\times10^{16}})^{2.5}} cm^{-3}}.
 \end{equation}
 The peak column density is 1.3\,\pdixb{23}\sqc, i.e. a visual extinction of \Avb 
 $\approx$\,140
  mag, compatible with the total absence of stars seen through the core even in the Spitzer MIR maps. 
It is also consistent with the dust emission at 850 \mum\ towards the same spot (which is a weak constraint given the large uncertainty
on the dust emissivity, \citealt{Demyk:2017bk,Demyk:2017eh}).\\
 We also apply this profile to the West side. Using a different density profile would locally change the abundance profile of the ions
 but would not impact the results for the central region of the core on first approximation.
\item  In the centre of cold cores, temperatures are usually in the 8 -- 12\,K range \citep{Caselli:2008gk}. We thus considered central temperatures of 8, 10, and 12\,K and imposed a 
temperature of 12\,K  in the external layers on the eastern side (the exact value has no impact in the 11-16\,K range).
\item The \ndhpb\jqtb emission peaks about 20\arcsec\ west of the PSC centre in a place that is warmed by the outflow impact. Former studies proposed temperatures between 
16 and 25K 
\citep{Loinard:2001im,Castets2001,Lis2002,Stark:2004hz,Lis:2016gx}, and we display results for the median value of 20K but also checked the 16 and 25\,K cases.
\item The para--\ddhpb spectra from $-14$\arcsec\ to $-28$\arcsec\ (and possibly $-35$\arcsec) in Fig. \ref{fig:figure3-1d2hph2dpfit} are slightly stronger than their symmetrical 
ones 
eastward, 
and since the 
position of the peak of \ndhpb emission is in the middle of this range, we set the temperature increase at -20\arcsec.
\item We divided the sphere in three parts: each time we ran a 1D spherical model but we extracted the emission from a fraction of the sphere only.
We stitched the three extracted portions together 
and convolved the output to 14\arcsec. This is graphically explained in Appendix \ref{modelbuilding}.
The spherical model is discretized into 20 layers of 1.14\,\,\pdixb{16}\,cm thickness. This corresponds to 5.4\arcsec\ thickness per layer 
for a distance of 141\,pc. The total size is 216\arcsec, which completely covers  the core. 
\end{itemize}

Though the warm part could be different in its density profile due to the possible compression from the outflow, if any (the outflow could be grazing the PSC and have a limited 
compression effect only), it is not possible to assess the actual profile at this stage until we 
have
 built a 3D model to explore that possibility. However the warm part is far enough from the horse--shoe central region and has no influence on the modelling
 of the cloud centre, as our tests have shown. Similarly, the impact of the temperature between 16 and 25 K was found to be negligible.

\begin{figure}
\centering
\includegraphics[width=1.0\linewidth]{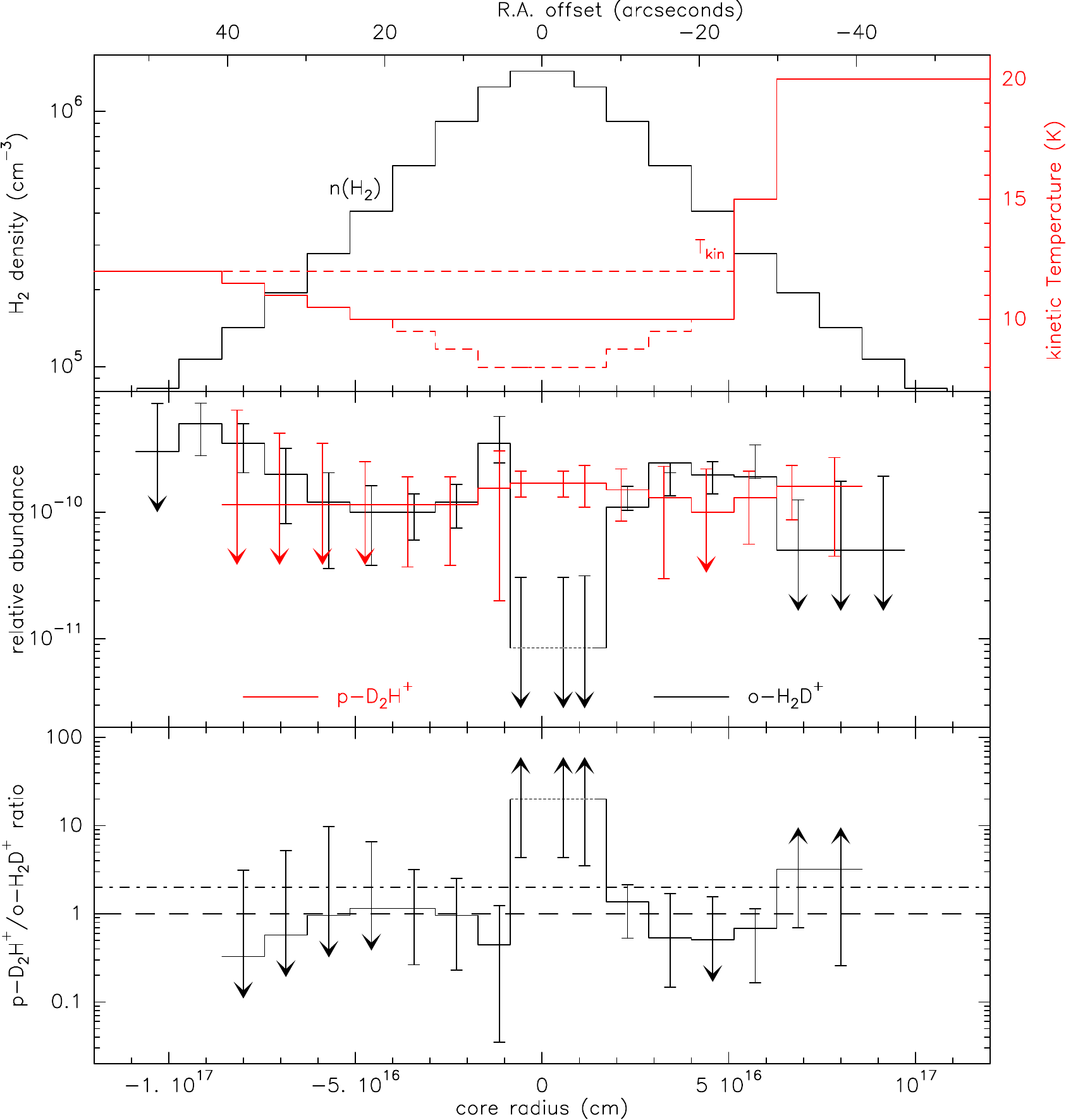}
\caption{{Pseudo-1D model profiles of the core. {\bf Upper box:}  n(H2) Plummer-like profile in black. In red,  temperature profile treated separately for the East (left) and the West 
(right) parts of the core, with three different minimum temperatures (10\,K in plain line, 8 and 12\,K in dashed lines). {\bf Middle box:} ortho--\hddpb (black) and para--\ddhpb 
(red) 
abundance profiles relative to \hdb and computed separately for each 
side for the 10K model, error bars are 1 $\sigma$ deviation from best 
model. Arrows mark upper limits. In particular, for ortho--\hddpb in the centre of the core, the dotted line represents the abundance to stay just below 1 $\sigma$ difference with 
the 
observations, and the upper error bar represents a difference of 2 $\sigma$ with the observations. 
 {\bf Lower box:} \label{key}para--\ddhp/ortho--\hddpb ratio for 
the 10K 
model. 
Error bars are computed by taking the opposite error bars of both species when available, otherwise lower or upper limits are indicated by arrows. The long dashed and dash-dotted 
lines  
mark a ratio of 1 and 2, respectively, 2 being the upper limit of the models of \citet{bovino_chemical_2021} at 10\,K.}}
\label{fig:figure4cdoh2dpc3pd2hpc1ratio2gris}
\end{figure}

\begin{figure}[t!]
\centering
\includegraphics[width=\linewidth]{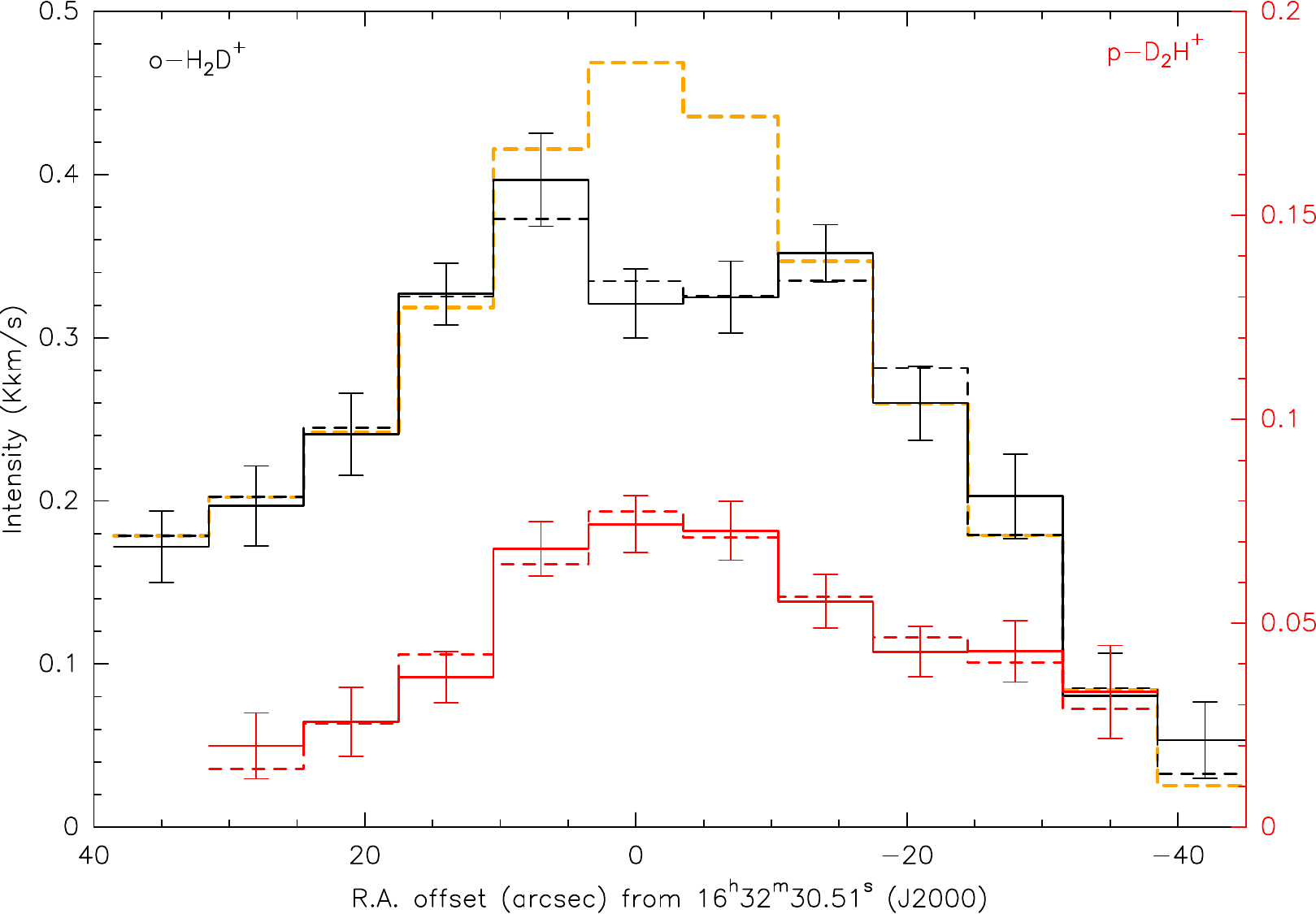}
\caption{Observed intensities along the horizontal cut (full lines with their error bar, corresponding to the spectra from Fig. \ref{fig:figure3-1d2hph2dpfit}) and corresponding 
model intensities (dashed lines). For ortho--\hddp, we present an additional  model (orange dashed line) for which the abundance is kept constant across the core. Para--\ddhpb 
(in red) observations and model have 
their own intensity axis, also in red, on the right side of the plot.}
\label{fig:figcoupeobsmodelintensityhighabun}
\end{figure}

The modelled density, temperature, and isotopologue abundance profiles are displayed in Fig. \ref{fig:figure4cdoh2dpc3pd2hpc1ratio2gris}. }
We adjusted the abundance,  non-thermal linewidth and radial velocity to fit the widths and strengths of the ortho--\hddpb and para--\ddhpb lines 
({see synthetic spectra in }Fig.\ref{fig:figure3-1d2hph2dpfit}). Linewidths show some inconsistencies as revealed by 
Gaussian fits to the observed spectra  (for more details, see Appendix \ref{app:linewidth}). 
{This pseudo-1D modelling is performed with a 1D Monte-Carlo code adapted from \citet{Bernes1979}'s radiative transfer program.} 
For the non-LTE modelling of the two ions, we used the inelastic collisional rate coefficients from \citet{Hugo:2009gq}.
The{ abundances }of both ions relative to \hdb are displayed in the {central} panel of Fig. \ref{fig:figure4cdoh2dpc3pd2hpc1ratio2gris} {while their ratio
is displayed in the bottom panel}. 

 The modelled lines are displayed 
in Fig. \ref{fig:figure3-1d2hph2dpfit} and the same fit is traced in terms of integrated intensity
in Fig. \ref{fig:figcoupeobsmodelintensityhighabun}, where the centre of the ortho--\hddpb horse--shoe shape is clearly visible. In  Fig. 
\ref{fig:figcoupeobsmodelintensityhighabun}, 
the observed 
intensities (full lines, with their individual error bars) are reproduced within 1 $\sigma$ by the 
model (dashed line, same colour). Because the density keeps increasing towards the centre, the ortho--\hddpb intensity drop needs a strong
abundance decrease to explain the decorrelation between the two (unlike para--\ddhp, the intensity of which keeps increasing). Since the para--\ddhp abundance is found to be
almost constant in the centre of the cloud (between 1.15 and 1.7\,\pdix{-10}, see below), we tested the case of a constant abundance for ortho--\hddp. The predicted intensity is 
shown in  Fig. 
\ref{fig:figcoupeobsmodelintensityhighabun} with a dashed orange line. 
Towards the central position, the difference reaches 7 $\sigma$ and a uniform abundance of ortho--\hddpb is clearly excluded.

\section{Discussion}
The \hddp/\ddhp contrast reported in Sect. \ref{Analyse} is expected from chemical evolution of the deuteration of the \htpb ion 
\citep{Roberts2003,Walmsley2004} that indicates that \htpb can be converted to \hddp, itself converted to \ddhpb and finally to \dtp, the latter being 
potentially the 
most abundant ion in the end. The effect is maximum at the peak density because the higher the density, the faster the chemical reactions, other parameters
being equal. This is the first time that we directly witness the intermediate step of this transformation (\htpb and \dtpb are not observable in 
dense 
clouds). 
This confirms that the 
original centre of the PSC is not coincident with the peak emission of species such as \ndhp, \nddp, and \dcop, {but 22\arcsec apart, }which differs from what is  usually 
found in 
other 
objects \citep[e.g.,][]{Tafalla2002, Pagani2007}. 
This must result from the temperature enhancement produced by the molecular outflow hitting the PSC.
{This heating possibly explains the similar extent of \hddpb and \ddhpb towards the working surface of the PSC, in the south-west direction. It may also explain the small 
secondary peak of para--\ddhpb as a result of the temperature increase compensating the drop of abundance, although further work is needed to reach a firm conclusion.}

{The column densities of both species derived from the modelling in Sect. \ref{sub:paraortho} are nearly equal towards the centre, N(ortho--\hddp) peaks at  2.1-- 
4.8\,\pdixb{13}\,\sqc\  (at 12\,K -- 8\,K), and 
N(para--\ddhp)  at 0.94 -- 4.2\,\pdixb{13}\,\sqc. The 
maximum  para--\ddhp/ortho--\hddpb column density ratio is 1.9$_{-1.2}^{+2.1}$, (at 10\,K, ranging from 1.1 at 12\,K to 3.1 at 8\,K)\footnote{{The peak column
densities are not coincidental.}}. However, the derived abundance profiles yield a 
stronger contrast 
in the inner shells as demonstrated in  Fig. \ref{fig:figcoupeobsmodelintensityhighabun}.
For the two most inner shells (total radius of 10.8\arcsec), the abundance of ortho--\hddpb can be set to zero while the line of sight emerging intensity computed by the model is still 
 too high (but less than 1 $\sigma$ away from the observations). Since we asymptotically reach a difference of +0.8 $\sigma$ for the central position by decreasing the
abundance by almost two orders of magnitude, we selected the abundance that provides exactly 1 $\sigma$ difference as the reference and computed the upper limit of this 
abundance by setting the error to 2 $\sigma$. There is no lower limit since we cannot make the model weaker than the observations for this central spectrum.
This is because the large ortho--\hddpb abundance needed to 
reproduce the intensity of the line towards the walls of the
horse--shoe cavity is also sufficient to reproduce the intensity of the line for the two sightlines across the horse--shoe hole, the same walls probably standing in front 
and in the back of the cavity, the observer having no privileged point of view 
(a situation similar to the case of \object{L1506C}, \citealt{Pagani2010}). 
{In the centre of the cavity we find a very high para--\ddhp/ortho--\hddp abundance ratio of $\sim$20. To check whether this ratio could be compatible with known models when 
taking into account the uncertainties of the line modelling, we computed a conservative lowest value of the uncertainty range of this ratio. }We estimated an upper limit to the 
abundance 
of ortho--\hddpb at 2 $\sigma$ above the observations  and found abundances of 1.8\,\pdixb{-11} at 12\,K, 
3.05\,\pdixb{-11} at 10\,K, and 4.0\,\pdixb{-11} at 8\,K. We also computed a lower abundance for para--\ddhpb by departing from the best fit abundance by $-$1$\sigma$
 (X[para--\ddhp] = 6.95\,\pdix{-11}, 1.3\,\pdix{-10}, and 3.3\,\pdixb{-10} at 12, 10, and 8K respectively).
The para--\ddhp/ortho--\hddpb ratio lowest limit ranges from 3.9 to 8.3 (at 12\,K -- 8\,K).
}

This para--\ddhp/ortho--\hddpb abundance ratio is high and beyond all predictions made by various models whether 0D
\citep{Pagani2009a, Parise:2011fk}, 1D \citep{pagani_ortho-h2_2013} or 3D 
\citep{bovino_3d_2019,bovino_chemical_2021} which reach values around 1--2 in the densest parts of the cores, {for a temperature of 10\,K}. 
\citet{Parise:2011fk} also found a  too high ratio when 
analysing the \object{H-MM1} prestellar core in the \object{LDN 1688} star forming region with a low temperature (7\,K) but could reconcile the  
model and observations at 12\,K.
{Here, the 12\,K value is possibly marginally consistent with the Bovino models if we interpolate by eye between their 10 and 15\,K cases \citep[their Fig. 
B.2]{bovino_chemical_2021}. However, clouds with \Av\,$\geq$ 100 mag. 
should be very cold in their centre, below 10\,K following the models of \citet{Zucconi:2001dk} and observations of \citet{Crapsi2007} and \citet{Pagani2007}.}
These observations coupled to the present {simple 1D modelling,{ if confirmed by a more accurate 3D modelling,}
either suggest }that current chemical models are not yet close enough to reality to trace correctly the 
deuteration evolution of the tri-hydrogen ion in the cold medium or that the reaction rate coefficients or inelastic collisional rate coefficients of the 
\htpb family with 
H$_2$ and HD from \citet{Hugo:2009gq}, which were computed semi--classically,  are not accurate enough and need to be improved. {We found that increasing the collision rate 
coefficients 
of para--\ddhp with H$_2$ by a 
factor 6 or increasing the ortho--\hddp + HD $\rightarrow$ para--\ddhp + para-H$_2$ reaction rate coefficient by a factor 3 would change the  para--\ddhp/ortho--\hddpb ratio 
by a 
factor 2. Other 
similar 
changes can also be 
considered.}

 \section{Conclusions}
 We present in this article the first map of para--\ddhpb in the interstellar
 medium, obtained with the APEX telescope toward the prestellar core
 \object{IRAS 16293E}. The main conclusions of this work are:   
 \begin{enumerate}
\item Maps of para--\ddhp, and to a lesser degree, of ortho--\hddpb reveal the position of the core centre before it was compressed and heated by at least one outflow coming 
from
 nearby YSOs. The new reference position for the core centre is {$\alpha_{J2000}$: 16$^h$32$^m$30.51$^s$, 
  $\delta_{J2000}$:  $-$24$\degr$28$^\prime$53.7$^{\prime\prime}$.}
\item We derived peak \cdsb of $\sim$ 1--3 \pdix{13} \sqcb for both ions {(at 10\,K).}
 \item We derived a conservative lower limit  of the para--\ddhp/ortho--\hddpb abundance ratio in the centre of the PSC  in the range 3.9--8.3 at 12 -- 8\,K {while the core 
 extinction 
 (\Av $\geq$ 100  mag) is indicative of a central temperature below 10\,K.}  {This ratio is higher than what 
 published chemical models predict for temperatures below or around 10\,K.}
 \item Temperature effects  induced by the outflow impact at the hotspot make line emission of most species bright. 
 Therefore an  analysis limited to the 
 hotspot could miss the real peak abundance of species, in particular ND$_3$ and other deuterated species, the emission of which  peaks away from 
 the ortho--\hddpb and 
 para--\ddhpb abundance peaks.
 \end{enumerate}
 {To confirm the present  model, }a thorough modelling of the core to disentangle density and temperature variations induced by the outflow impact needs to take 
 into account all 
 available information
 (line emissions with several transitions per species, dust emission and absorption lower limit) in a 3D model, {though the main difficulty of this model will be to constrain the line 
 of sight distribution of all parameters}. 
 
The high para--\ddhp/ortho--\hddpb ratio measured in \object{IRAS 16293E}  {could} suggest, {if confirmed}, that the inelastic collisions or reaction rate coefficients of the 
\htpb 
isotopologues  need to be revised 
or that the chemical models need to be refined in order to improve our understanding of the deuteration process in star forming regions.
 This might have an impact on the deuteration chemical clock used to study many 
 star forming regions.
 
 \begin{acknowledgements}
Based on observations with the Atacama Pathfinder EXperiment (APEX) telescope. At the time of the first observations presented in this paper, APEX was a collaboration between 
the Max Planck Institute for Radio Astronomy, the European Southern Observatory, and the Onsala Space Observatory.
This research has made use of NASA’s Astrophysics Data System Abstract Service.
It  has also made use of observations from Spitzer Space Telescope and from the NASA/IPAC Infrared Science Archive, which are operated by the Jet 
Propulsion 
Laboratory (JPL) and the California Institute of Technology under contract with NASA.
This research makes a large use of the CDS (Strasbourg, France) services, especially Aladin \citep{Bonnarel2000}, Simbad and Vizier \citet{Ochsenbein2000}.
We are indebted to C. de Breuck for the observations in the ESO time and to all the observers at APEX. We thank P.F. Goldsmith for useful comments and M. Tafalla for his
helpful advices and editorial work.

 \end{acknowledgements}
 \bibliographystyle{aa}
 
 \bibliography{references,zotero,papers}
 
\begin{appendix}
\section{Reference positions for the APEX para--\ddhpb observations}
\begin{table*}
\caption{{Reference positions for the APEX observations}}
\label{tab:refpos}
\centering
\begin{tabular}{ccccc}
\hline\hline
Project &date&observing mode&R.A.(J2000) or offset\tablefootmark{a}&Dec (J2000) or offset\tablefootmark{a}\\
\hline
M-085.F-0016-2010&07/09/2010&position switch&$-$300\arcsec& 0\\
E-0105.C-0251A-2020&03/05/2021&wobbler switch&±200\arcsec\ (H)&0\\
E-0105.C-0251A-2020&28--29/06/2021&OTF&16h32m40s&$-$24\degr28\arcmin30\arcsec\\
M-109-.F-9501C-2022&24/03--04/04/2022&OTF &16h32m40s&$-$24\degr28\arcmin30\arcsec\\
\hline
\end{tabular}
\tablefoot{
\tablefoottext{a}{{Offsets are with respect to equatorial coordinates except if marked with (H) for horizontal wobbler switching}}
}
\end{table*}

{Different reference positions were used for the various projects with APEX towards \object{IRAS 16293E}.
We list them in Table \ref{tab:refpos}}

\section{Model building} \label{modelbuilding}
\begin{figure}[h]
\centering
\includegraphics[width=\linewidth]{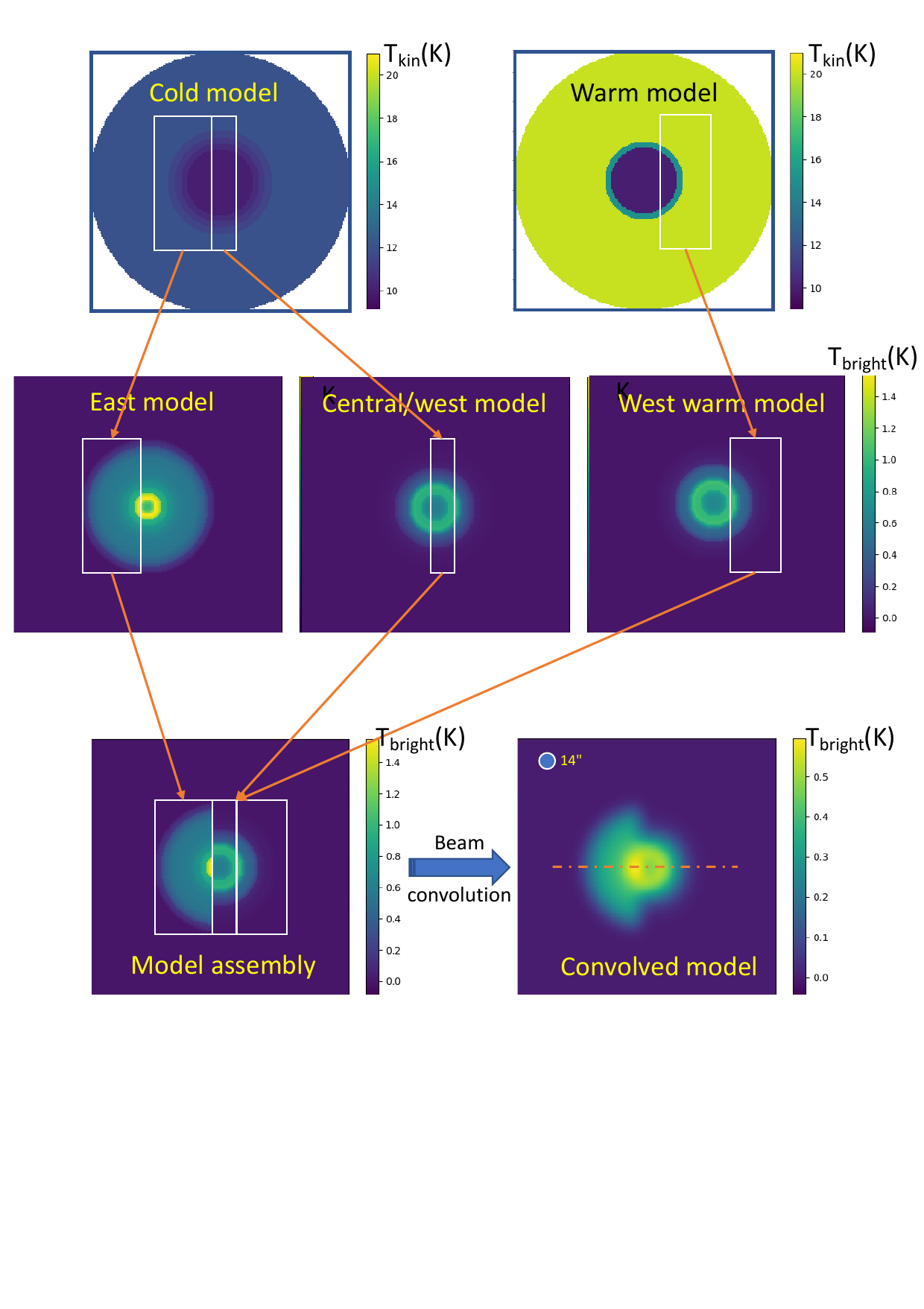}
\vspace{-3.3cm}
\caption{Steps to build an asymmetrical model from 1D spherical models. Based on a single density profile (eq. \ref{plummer}), we compute two models with different temperature 
profiles in 
the outer parts (either 12 or 20 K, top row, see Fig. \ref{fig:figure4cdoh2dpc3pd2hpc1ratio2gris}). We apply two different abundance profiles (central row) for each ion for the East 
and West parts, the West part is computed for both 
temperature 
profiles. We extract the parts corresponding to each region of interest (sampled with a 1\arcsec\ resolution) and we assemble them in a cube (RA, Dec, Frequency, bottom row left). 
Finally, we convolve the cube with a 14\arcsec\ half-power beamwidth Gaussian beam to compare to the observations (bottom row right). We extract the spectra along the dash-dot 
line 
and 
compare them to the observations 
(see Fig. \ref{fig:figure3-1d2hph2dpfit}).}
\label{fig:constructionmodele}
\end{figure}

The model is based on a single density profile (eq. \ref{plummer}) and we computed the emission on a 1\arcsec\ step basis (model diameter of 216\arcsec) for various temperature 
and abundance profiles.
\begin{enumerate}
\item for positive offsets (East side) we used the pristine model and extracted the emission from the half sphere towards the East.
\item for offsets between 0 and -15\arcsec, we used the 
same model but on 
the West side with a different abundance profile except in the central shell which is forcibly common to both sides, being unresolved,
(the centre being common to the East and West sides, we technically use only the West side and therefore the East stops at 5\arcsec, and the West goes from $-$15 to $+$5\arcsec )
\item for the external heated part, selected beyond -15\arcsec\ from the centre, we used a 1D model with warm outer layers beyond a radius of 20\arcsec\ in the Western part. 
\end{enumerate} 
For the ortho--\hddpb case shown in Fig. \ref{fig:constructionmodele}, the strong dissymmetry is due to the fact that the emission is extended to the East and not as much to the 
West. 
The ortho--\hddpb abundance profiles (Fig. \ref{fig:figure4cdoh2dpc3pd2hpc1ratio2gris}, central panel) reflect this difference by decreasing beyond $-$30\arcsec\ on the Western 
side, while it keeps increasing beyond $+$30\arcsec\ on the Eastern side.
One should note that the warm layers beyond -20\arcsec\ contribute to positions 
 closer to the cloud centre, because being a 1D onion model, they are present on the line of sight of these closer positions in front and behind. The line of sight crosses a cold 
 layer inserted between two hot layers. This could be justified by the outflow interacting with the surface of the PSC rather than ploughing its way to the centre, since the 
 ortho--\hddpb and para--\ddhpb lines do not seem to be kinematically perturbed in this warmer region. This problem will be 
 examined in detail in a future work.
The model is not representative of the emission towards the North or the South of the PSC.

\section{Linewidth and Gaussian fits of the \ddhpb APEX and \hddpb JCMT spectra}\label{app:linewidth}

\begin{figure*}[t!]
\centering
\includegraphics[width=\linewidth]{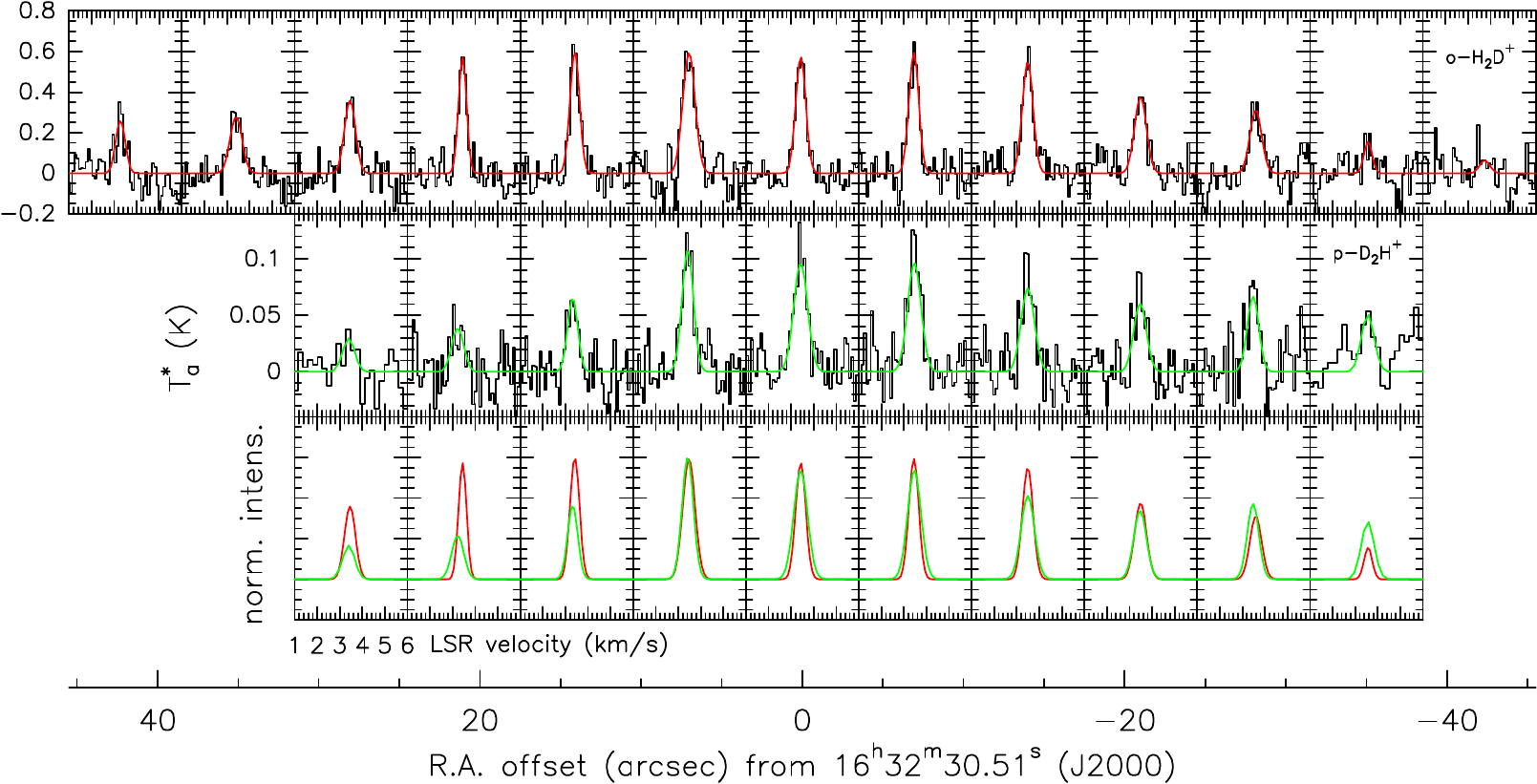}
\caption{{{
JCMT spectra of  ortho--\hddpb\jkk:1\down{10}--1\down{11} (top row,
velocity axis from 1 to 6\,\kmps), and APEX spectra of para--\ddhpb\jkk:1\down{10}--1\down{01} (middle row, velocity axis from 1 to 6\,\kmps) across
the cloud (the position of the cut is indicated in Fig. \ref{fig:figure2H2DpD2Hp}). Gaussian fits are overlaid in red and green, respectively. The same Gaussian fits with same 
colour code are reproduced 
in the bottom row, after normalization to the peak temperature for the +7\arcsec spectrum, for a better comparison of the ratio variation and linewidth differences between the two 
isotopologues}.}}
\label{fig:figAnn1GaussFit}
\end{figure*}

{
The lines of both ions are optically thin (peak opacity of 0.6 and 0.7 for para--\ddhpb and ortho--\hddp, resp.), 
and in first approximation consistent with a Gaussian shape. We have therefore adjusted a
Gaussian to each of the spectra. These Gaussian fits are displayed 
in Fig. \ref{fig:figAnn1GaussFit}. To adjust the Monte-Carlo model, we modified the isotopologue abundance profile across the core to minimize 
the integrated intensity difference between the observations and the modelled line. Since the lines are optically thin, the exact 
shape of the line has little impact on its integrated intensity but of course, we take care that the linewidth be compatible with the observations. However, from the Gaussian fit, 
we found that the linewidth of the two 
isotopologues are not equal. At the edges, this is probably a noise limitation (e.g. the 21\arcsec\ para--\ddhpb spectrum), but in the centre, the para--\ddhp spectra at offsets 
0, $-$7\arcsec, and $-$14\arcsec\ are clearly larger than the ortho--\hddpb ones, e.g., for the 0 position: 
\begin{itemize}
\item $\delta$V(para--\ddhp) = 0.658 $\pm$0.074\,\kmps
\item $\delta$V(ortho--\hddp) = 0.524 $\pm$0.037\,\kmps
\end{itemize}
}

{\hddpb being lighter than \ddhp, one would expect the former to display lines slightly wider than the latter for similar thermal turbulence or macroscopic movements.
The reverse could indicate that the core inside the horse--shoe, depleted in \hddpb (see Sect. 3), has a signature of collapse of its own traced by \ddhpb alone,
but we did not manage to reproduce this difference with our model. The volume is probably too small to dominate the linewidth formation without perturbing the \hddpb 
linewidth too.
The other possibility is that this difference be mostly due to noise (1.8 $\sigma$ difference) and we settled the model linewidth to be intermediate between these two values, making 
the \ddhpb
modelled line slightly stronger than the Gaussian fit but narrower, and conversely for \hddp.
}
 \end{appendix}
 \end{document}